# Condition of Multibubble Sonofusion and Proposal of Experimental Setup

Masanori SATO[1], Hideo SUGAI[2], Tatsuo ISHIJIMA[2],

Hirotaka HOTTA[2], Masahiro TAKEICHI[2], Nagaya OKADA[1]

[1]Honda Electronics Co., Ltd., 20 Oyamazuka, Oiwa-cho, Toyohashi, Aichi 441-3193, Japan

E-mail:msato@honda-el.co.jp

[2]Nagoya University, Furo-cho, Chikusa-ku, Nagoya, Aichi 464-8603, Japan

**Abstract**: We consider that multibubble sonoluminescence (MBSL) sonofusion is necessary for the industrial use of sonofusion. In 2002, Taleyarkhan et al. [Science, **295**, 1868, (2002)] reported neutron radiation from single-bubble sonoluminescence (SBSL), thereafter sonofusion has been discussed. However, SBSL, as described in the literature, is only one bubble, then the energy of sonofusion has a strict limitation for industrial use, that is, very little production of energy. MBSL is a cluster of bubbles; therefore, if sonofusion occurs in MBSL, a large amount of energy can be obtained. We assume the difference between SBSL and MBSL is temperature, thus we propose experimental setup of sonofusion, in which sonoplasma raises the temperature of MBSL.

1. Introduction

   Sonofusion, that is, a type of cold fusion, has attracted much attention, particularly in the USA. Sonofusion has been discussed since the patent of Flynn [1] in 1978, the importance of the patent of which was discussed by Crum [2]. Putterman [3] issued a similar patent in 1994 and since then there have been many patent issues and applications.

   Recently, Taleyarkhan et al. [4, 5] have reported neutron detection from SBSL, attracting attention. Saltmarsh et al. [6] carried out the experiment using the experimental setups of Taleyarkhan's group and reported that there is no statistical evidence of neutron radiation from SBSL. Taleyarkhan made a counterargument against Saltmarsh's conclusion using Saltmarsh's data, that is, there is neutron radiation from SBSL. In 2005, Xu et al. [7] reported neutron radiation from SBSL, which confirmed Taleyarkhan's experimental data.

   Experiments under other conditions, that is, without preradiation of neutrons to SBSL were carried out. Neutrons were not detected from laser-induced cavitation in heavy water by Geisler et al. [8]. Didenko et al. [9] reported that no nuclear fusion occurs through cavitation because cavitation dissipates the energy during collapse.

In the experiments by Taleyarkhan et al. [4, 5] and Xu et al. [7], the preradiation of neutrons for the generation of SBSL nuclei was performed. The preradiation of neutrons has the possibility of stimulating benzene, after which SBSL induces sonofusion. Taleyarkhan et al. [10] reported nuclear emissions during neutron-seeded acoustic bubble cavitation. Xu et al. [7] reported that they did not detect neutrons from SBSL without pre- or simultaneous-radiation of neutrons. Therefore, pre- or simultaneous-radiation of neutrons is indispensable for sonofusion.

Cavitation and sonoluminescence have been discussed from the viewpoint of plasma. Flannigan [11] reported that cavitation bubbles are in a plasma state. Sonoplasma, that is, cavitation induced discharge plasma [12] and microwave plasma [13], was utilized for producing amorphous carbon and nano-carbon particles. Cavitation and sonoluminescence are closely related to plasma.

Yasui et al. [14] reported the sonoluminescence totally. In their paper, rare gases (argon and xenon) play an important roll in the brightness of sonoluminescence, that is, under the condition of a rare-gas-saturated liquid, the temperature of bubbles increases and intense sonoluminescence can be observed. The mechanisms of SBSL and MBSL are also discussed. In MBSL, where acoustic amplitude is relatively small, that is 2 bar at 20 kHz, light emissions from plasma inside collapsing bubbles are dominant as in the case of SBSL. At a relatively large acoustic amplitude, 3 bar at 20 kHz, the chemiluminescence of OH radicals is dominant as the mechanism of light emissions. The temperature of SBSL is above 10,000 K, and that of MBSL is below 10,000 K. Sato et al. [15] reported that cavitation bubbles are acoustic waves on the bubble surface and accumulate acoustic energy during several periods. SBSL flashes synchronized to the driving frequency, but MBSL should flash after accumulating acoustic energy for several periods.

In this report, we propose an experimental setup for MBSL sonofusion.

2. Condition of multibubble sonofusion

The difference between the two groups who detected and did not detect neutrons from SBSL is a result of pre- or simultaneous-radiation of neutrons to SBSL. Taleyarkhan et al. [4, 5] and Xu et al. [7] used pre- or simultaneous-radiation of neutrons, however other groups, which did not detect neutron radiation from SBSL, did not use pre- or simultaneous-radiation of neutrons. Sonofusion seems to require pre- or simultaneous-radiation of neutrons.

In the industrial application of sonofusion, there is a marked difference in the condition between SBSL and MBSL. The condition of MBSL makes the industrial application of sonofusion easy. SBSL, reported in the literature, is only a bubble, then the region of sonofusion is limited to very small spot, thus sonofusion generates very little energy. MBSL is a phenomenon where a large number of bubbles emit light, therefore the efficiency of sonochemical reaction is much higher in MBSL than in SBSL. The difference between SBSL and MBSL is the temperature during collapse.

Yasui et al. [14] reported that the temperature of SBSL is above 10,000 K, and that of MBSL is below 10,000 K. If the temperature of SBSL is required, we can satisfy this condition using sonoplasma, and the temperature of the plasma is estimated to be above 10,000 K. There are no experimental data of MBSL with pre- or simultaneous-radiation of neutrons. At this stage, we cannot verify that a temperature of 10,000 K in SBSL is required for sonofusion.

Nuclear fusion requires a temperature of 1,000,000 K, that is, the temperature of SBSL or plasma of approximately 10,000 K is insufficient. Taleyarkhan et al. [4, 5, 10] reported that the preradiation of neutrons induces a meta-stable state in deuterium atoms.

3. Proposal of experimental setup

Several experimental setups of sonoplasma have been reported [12, 13]. We are now preparing the experimental setup shown in **Fig. 1**.

We are carrying out the experiment on sonoplasma in benzene. Ultrasonic waves (20 kHz, ~1 kW) and pulsed microwaves (2.45 GHz, ~1.8 kW) are simultaneously irradiated in benzene with argon gas bubbling. In benzene, sonoplasma, that is, many flashing bright spots, can be observed as shown in **Fig. 2**. In water, sonoluminescence intensified by argon was detected below the ultrasonic horn. Sonoluminescence was intensified by argon, which could be observed by the naked eye.

**Figure 3** shows our proposed experimental setup of sonofusion. As discussed in section 2, a neutron emitter and a neutron counter are set to detect emitted neutron from MBSL. We can obtain thermal energy using a cooling pipe in the setup of sonoplasma. We are preparing deuterated acetone and a neutron emitter and we will be starting the experiment on sonofusion.

4. Conclusion

In this report, we proposed an experimental setup for MBSL sonofusion. Experimental setup of sonoplasma has already been prepared. In the future, we will prepare: 1) deuterated benzene or acetone, 2) a neutron emitter, and 3) a neutron detector, and we will start our experiment on MBSL sonofusion.

References

1) H. Flynn, US 4333796, published 1982.
2) A. Crum, "Acoustically induced cavitation fusion", J. Acous. Soc. Am., **103**, 3012, (1998).
3) S. Putterman, B. Barber, R. Hiller, R. Maire, J. Lofstedt, US 5659173, published 1997.
4) R. Taleyarkhan, C. West, J. Cho, R. Lahey Jr., R. Nigmatulin, and R. Block, Science, **295**, 1868, (2002).
5) R. Taleyarkhan, J. Cho, C. West, R. Lahey Jr., R. Nigmatulin, and R. Block, Phys. Rev. E **69**,

036109 (2004).

6) M. Saltmarsh and D. Shapira, Science, **297**, 1603, (2002).

7) Y. Xu and A. Butt, "Confirmatory experiments for nuclear emissions during acoustic cavitation", Nuclear Engineering and Design, **235** (2005) 1317-1324.

8) R. Geisler, W.-D. Schmidt-Ott, T. Kurz and W. Lauterborn, Europhys. Lett., **66**, 435, (2004).

9) Y. Didenko, and K. Suslick, "The energy efficiency of formation of photons, radicals and ions during single-bubble cavitation", Nature, **394**, 418, (2002).

10) R. Taleyarkhan, C. West, J. Cho, R. Lahey Jr., R. Block, R. Nigmatulin, J. Acous. Soc. Am., **112**, 2269, (2002).

11) D. Flannigan, and K. Suslick, "Plasma formation and temperature measurement during single-bubble cavitation", Nature, **434**, 52, (2005).

12) R. Sergiienko, E. Shibata, H. Suwa, T. Nakamura, Z. Akase, Y. Murakami and D. Shindo "Synthesis of amorphous carbon nanoparticles and carbon encapsulated metal nanoparticles in liquid benzene by an electric plasma discharge in ultrasonic cavitation field", Ultrasonics Sonochemistry, (2005), in press.

13) S. Nomura, and H. Toyota, "Sonoplasma generated by a combination of ultrasonic waves microwave irradiation", Appl. Phys. Lett., **83**, 4503, (2003).

14) K. Yasui, T. Tsuziuti, M. Sivakumar and Y. Iida, "Sonoluminescence", Appl. Spect. Rev. **39**, 399, (2004).

15) M. Sato, N. Shibuya, N. Okada, T. Tou and T. Fujii, Phys. Rev. E **65**, 046302, (2002).

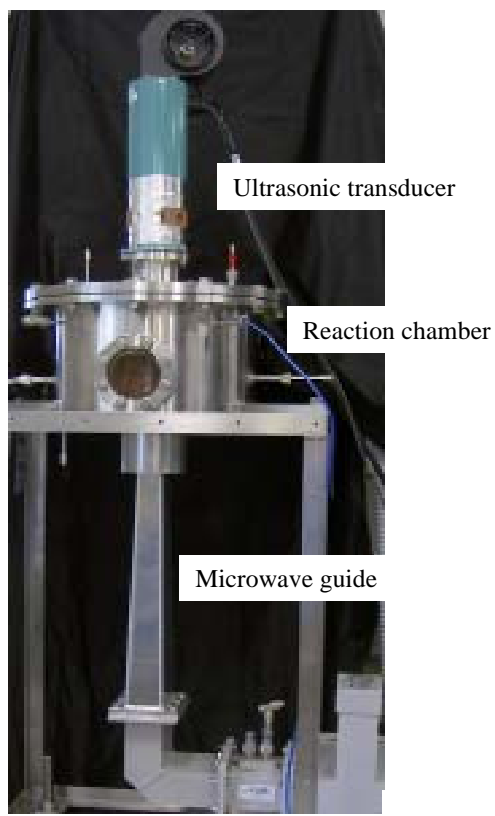

**Fig. 1** Experimental setup of sonoplasma.

The ultrasonic transducer is set at the top and the microwave guide is set at the bottom. In the reaction chamber, the simultaneous irradiation of ultrasonic waves and microwaves is carried out. We can observe the sonoplasma in the chamber through windows.

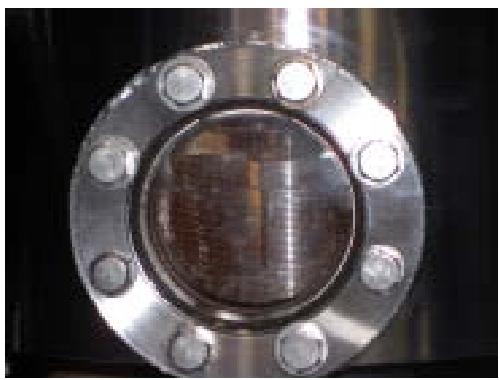

(a)

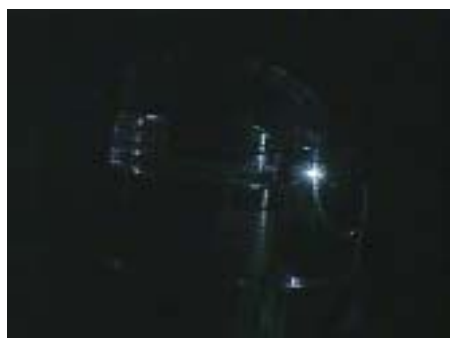

(b)

**Fig. 2** (a) Window of reaction chamber and (b) sonoplasma in argon-gas-bubbled benzene with helium flow.

Benzene in a beaker set in the chamber is argon-gas-bubbled, and helium flow is used to remove air in the chamber. Many flashing bright spots can be observed through the windows during the simultaneous irradiation of ultrasonic waves and pulsed microwaves. Flashing cannot be observed with only pulsed microwave irradiation.

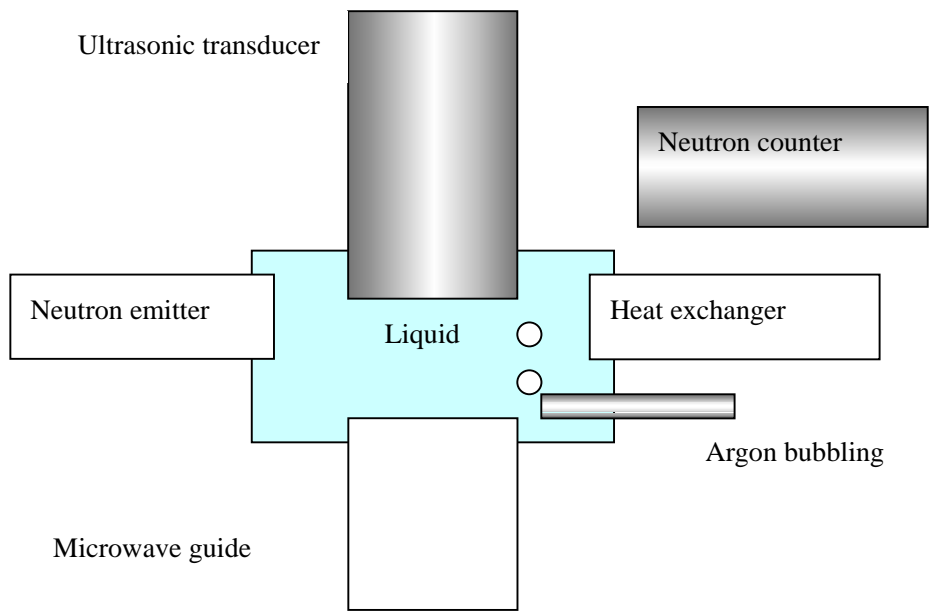

**Fig. 3**  Proposal of experimental setup of multibubble sonofusion

We are now preparing liquid deuterated acetone and a neutron emitter, and in the future we will start our experiment on sonofusion.